\documentclass[twocolumn,showpacs,preprintnumbers,amsmath,amssymb,bf]{revtex4}
\usepackage[caption=false]{subfig}
\usepackage{graphicx}

\begin{document}

\title{Constraints on exotic spin-dependent interactions between matter and antimatter from antiprotonic helium spectroscopy}

\author{Filip Ficek$^1$}
 \email{filip.ficek@uj.edu.pl}
\author{Pavel Fadeev$^{2,3}$}
\author{Victor V. Flambaum$^{2,4}$}
\author{Derek F. Jackson Kimball$^5$}
\author{Mikhail G. Kozlov$^{6,7}$}
\author{Yevgeny V. Stadnik$^{2}$}
\author{Dmitry Budker$^{2,8,9}$}
\affiliation{
$^1$ Institute of Physics, Jagiellonian University, \L{}ojasiewicza 11, 30-348 Krak\'{o}w, Poland\\
$^2$ Helmholtz Institute Mainz, Johannes Gutenberg University, 55099 Mainz, Germany\\
$^3$ Ludwig-Maximilians-Universit\"{a}t, M\"{u}nch\"{e}n, Fakult\"{e}t f\"{u}r Physik, Arnold Sommerfeld Center for Theoretical Physics, 80333 M\"{u}nchen, Germany\\
$^4$ School of Physics, University of New South Wales, Sydney, New South Wales 2052, Australia\\
$^5$ Department of Physics, California State University - East Bay, Hayward, California 94542-3084, USA\\
$^6$ Petersburg Nuclear Physics Institute of NRC ``Kurchatov Institute", Gatchina 188300, Russia\\
$^7$ St. Petersburg Electrotechnical University LETI€, Prof. Popov Str. 5, 197376 St. Petersburg, Russia\\
$^8$ Department of Physics, University of California at Berkeley, Berkeley, California 94720-7300, USA\\
$^9$ Nuclear Science Division, Lawrence Berkeley National Laboratory, Berkeley, California 94720, USA
}
\date{\today}

\begin{abstract}
Heretofore undiscovered spin-0 or spin-1 bosons can mediate exotic spin-dependent interactions between standard-model particles. Here we carry out the first search for semileptonic spin-dependent interactions between matter and antimatter. We compare theoretical calculations and spectroscopic measurements of the hyperfine structure of antiprotonic helium to constrain exotic spin- and velocity-dependent interactions between electrons and antiprotons.

\end{abstract}


\maketitle

\section{Introduction}
Antiprotonic helium (He$^{+}\overline{\text{p}}$) is a helium atom where one of the electrons is replaced with an antiproton. Antiprotonic helium, being a relatively simple matter-antimatter bound state, can provide insight into possible exotic matter-antimatter interactions. Since the first observations of relatively long-lived (lifetimes of the order of microseconds) antiprotonic helium atoms in 1991 \cite{Iwa91}, there have been significant developments in experimental techniques. The latest achievements include determining the antiproton magnetic moment \cite{Pas09, Nag17, Smo17}, resolving the hyperfine structure of $^3$He$^{+}\overline{\text{p}}$ \cite{Fri11} and precise measurements of the antiproton-to-electron mass ratio \cite{Hor16}. Furthermore, theoretical calculations of transition energies in antiprotonic helium have reached agreement with experiment at a level of one part in $10^{9}$ or better in many cases \cite{Bak98,Kor01,Kor06,Kor08,Hu16}. An extensive summary of research on antiprotonic helium prior to 2002 can be found in Ref.~\cite{Yam02}.

The principal focus of antimatter research to date has been on tests of CPT invariance \cite{Hay07}, for example, by measuring the properties of the antiproton \cite{Pas09,Nag17, Smo17,Hor16}, and on constraining Yukawa-type spin-independent forces \cite{Bor94,Sal14}. In this work, we show that one can also search for exotic spin-dependent interactions between matter and antimatter from precise measurements and QED-based calculations of antiprotonic helium.

Spin-dependent interactions \cite{Moo84,Dob06} appear in theories including ``new", i.e., so far undiscovered bosons such as axions \cite{Wei78,Wil78,Din81,Shi80,Kim79,Zhi80}, familons \cite{Wil82,Gel83}, majorons \cite{Gel81,Chi81}, arions \cite{Ans82}, new spin-0 or spin-1 gravitons \cite{Sch79,Nev80,Nev82,Car94}, Kaluza-Klein zero modes in string theory \cite{Svr06}, paraphotons \cite{Oku82,Hol86,Dob05}, and new $Z'$ bosons \cite{Bou83,App03,Dzu17}. These new bosons are introduced to solve problems such as the nature of dark matter \cite{Ber05} and dark energy \cite{Fri03,Fla09}, the strong-CP problem \cite{Moo84}, and the hierarchy problem \cite{Gra15}.

The most commonly employed framework for the purpose of comparing different experimental searches for exotic spin-dependent interactions is that introduced in Ref.~\cite{Moo84} to describe long-range spin-dependent potentials associated with the axion and later extended in Ref.~\cite{Dob06} to encompass long-range potentials associated with any generic spin-0 or spin-1 boson. Some issues related to the velocity-dependent potentials presented in Ref.~\cite{Dob06} were pointed out in Ref.~\cite{Fic17} and are resolved in Ref.~\cite{FadIP}. The spin-dependent potentials enumerated in Refs.~\cite{Dob06,FadIP} are characterized by dimensionless coupling constants that specify the strength of the interaction between various particles and a characteristic range parameter $\lambdabar$ for the interaction associated with the reduced Compton wavelength of the new boson of mass $m_0$, $\lambdabar = \hbar/(m_0 c)$ where $\hbar$ is the reduced Planck's constant and $c$ is the speed of light. Depending on the nature of the new interaction, different particles will generally have different coupling constants.

To date, the constraints on exotic spin-dependent interactions between matter and antimatter have concerned leptonic interactions and are derived from hydrogenlike atoms: positronium \cite{Les14,Kot15,Yev17} and muonium \cite{Yev17,Kar10,Kar11}. In the following, we constrain spin-dependent interactions between an electron and an antiproton (a semileptonic interaction). We do this in a similar manner to Ref. \cite{Fic17}, by comparing experimental results for the hyperfine structure of $^4$He$^{+} \overline{\text{p}}$ \cite{Pas09} and QED-based calculations \cite{KorBak01} and using our calculated expectation values of spin-dependent potentials.

The structure of this paper is as follows. We begin by constructing approximate wavefunctions describing the antiprotonic helium atom. Then we present the relevant exotic potentials. Finally, we use first-order perturbation theory on the aforementioned wavefunctions and potentials to obtain constraints on the interaction parameters of interest.

\section{Antiprotonic helium wavefunctions}
\begin{figure}
\includegraphics[width=0.4\textwidth]{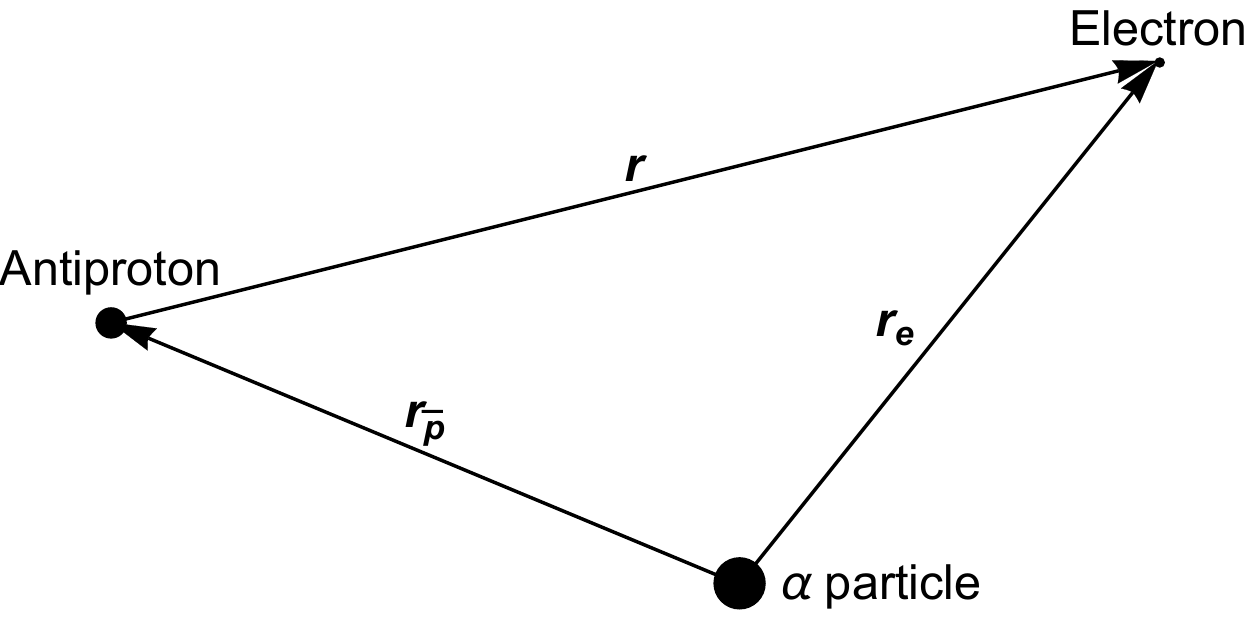}\\
\caption{\label{fig:phe} Schematic diagram of the antiprotonic helium atom. The nucleus is an alpha particle.}
\end{figure}

Since the electron mass $m_e$ is much smaller than the nuclear (alpha particle) and antiproton masses, $m_\textrm{nucl}$ and $m_{\overline{p}}$, respectively, the approximate Hamiltonian describing antiprotonic helium has the form (derived in Appendix \ref{sec:wave}):
\begin{align} 
\hat{H}=\left(-\frac{\hbar^2}{2 \mu_{\overline{p}}} \nabla_{\overline{p}}^2-\frac{2e^2}{|\mathbf{r}_{\overline{p}}|}\right)+\left(-\frac{\hbar^2}{2 m_e}\nabla_e^2-\frac{2e^2}{|\mathbf{r}_e|}\right)+\frac{e^2}{|\mathbf{r}_{\overline{p}}-\mathbf{r}_e |},
\label{eqn:hamiltonian0}
\end{align}
where $e$ is the elementary charge, $\mu_{\overline{p}}=m_\textrm{nucl} m_{\overline{p}}/(m_\textrm{nucl}+m_{\overline{p}})$ is the reduced mass of the antiproton, $\mathbf{r}_{\overline{p}}$ and $\mathbf{r}_e$ are the position vectors of the electron and antiproton with respect to the nucleus (Fig.\ \ref{fig:phe}), and $\nabla_{\overline{p}}$ and $\nabla_e$ are Laplacians in the coordinates $\mathbf{r}_{\overline{p}}$ and $\mathbf{r}_e$.

The strength of any hypothetical exotic spin-dependent interaction between two charged particles is orders-of-magnitude smaller than their electromagnetic interaction. Based on this, a high-precision calculation of the perturbation due to exotic effects is not required and it is sufficient to calculate the exotic contributions to first order in perturbation theory. For these calculations, a relatively simple form of the approximate wavefunctions of the antiproton and electron may be assumed. In the following, we focus on antiprotonic helium  with the antiproton in the $(n,l)=(37, 35)$ state and the electron in the $(1,0)$ state (where the first number in an ordered pair is the principal quantum number, and the second one is the orbital angular momentum), since there are both relatively precise experimental data and theoretical calculations available for this system \cite{Pas09, KorBak01}. As explained in Appendix \ref{sec:wave}, we use the approximate spatial wavefunction
\begin{align} 
\Psi_{\widetilde{m}_{\overline{p}}}(\mathbf{r}_{\overline{p}},\mathbf{r}_e)=&\frac{1}{\sqrt{1-\beta^2}} \left[\psi^{(\overline{p})}_{37,35,\widetilde{m}_{\overline{p}}}(\mathbf{r}_{\overline{p}}) \psi^{(e)}_{1,0,0}(\mathbf{r}_e)\right.\nonumber\\
&\left.-\beta \psi^{(\overline{p})}_{36,35,\widetilde{m}_{\overline{p}}}(\mathbf{r}_{\overline{p}}) \psi^{(e)}_{1,0,0}(\mathbf{r}_e) \right],
\label{eqn:sol1}
\end{align} 
where $\beta$ is a numerical constant and $\psi^{(a)}_{n,l,\widetilde{m}}$ is a generalised hydrogen-like atom wavefunction \cite{Gri} for a particle $a$ with principal quantum number $n$, orbital angular quantum number $l$, and magnetic quantum number $\widetilde{m}$:
\begin{align} 
\psi^{(a)}_{n,l,\widetilde{m}}(r,\theta,\phi)=\sqrt{\frac{4 (Z^{(a)}_n)^3\mu^3 (n-l-1)!}{n^4 (n+l)!}} \left(\frac{2Z^{(a)}_n \mu^{(a)} r}{n}\right)^l \nonumber\\
\times e^{-\frac{Z^{(a)}_n \mu^{(a)} r}{n}}L^{2l+1}_{n-l-1} \left(\frac{2Z^{(a)}_n \mu^{(a)} r}{n}\right) Y_{l}^{\widetilde{m}}(\theta,\phi).
\label{eqn:sol2}
\end{align} 
In formula (\ref{eqn:sol2}), $\mu^{(a)}$ denotes the reduced mass of particle $a$, ($\mu^{(\overline{p})}=\mu_{\overline{p}}$), $\mu^{(e)}\approx m_e$, $Z^{(a)}$ is the effective charge seen by particle $a$ in a state with principal quantum number $n$, $L^{2l+1}_{n-l-1}$ is the generalised Laguerre polynomial, and $Y_{l}^{m}$ is the spherical harmonic function. The parameters $\beta$ and $Z^{(a)}_n$ are derived in Appendix \ref{sec:wave} using the variational method. In Appendix \ref{sec:check}, we justify the use of this method in this case.

\begin{figure}
\includegraphics[width=0.4\textwidth]{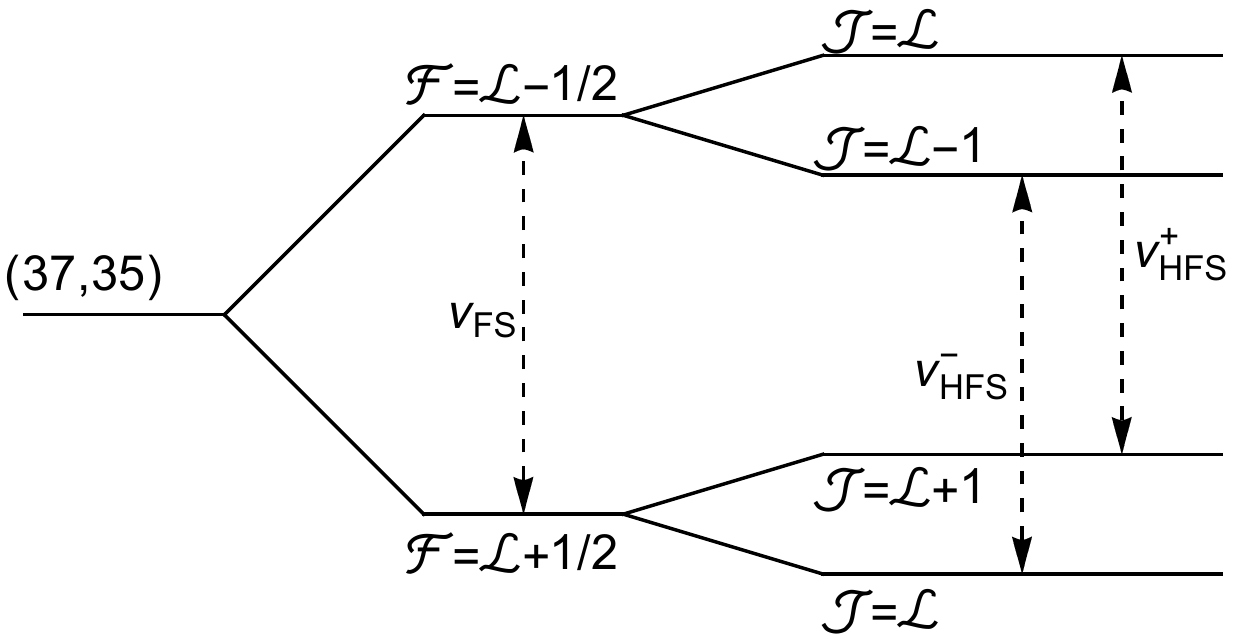}\\
\caption{\label{fig:hfs}Hyperfine structure of the $(n,l)=(37,35)$ state of an antiprotonic helium atom. The transitions denoted by $\nu_{\text{HFS}}^\pm$ were investigated in Ref.\ \cite{Pas09}.}
\end{figure}

To get a full approximate wavefunction of the considered system, we need to add the spinor component to the spatial wavefunction (\ref{eqn:sol1}). In the $(37,35)$ state, the total orbital angular momentum of the atom is $\mathcal{L}=35$. Let us then denote by $|35,\widetilde{m}_\mathcal{L}\rangle$ the vector corresponding to the spatial wavefunction $\Psi_{\widetilde{m}_{\overline{p}}}$. The interaction between the orbital motion and the electron spin is the strongest among the angular-momentum-dependent interactions in antiprotonic helium \cite{Bak98}, so we first add the orbital angular momentum to the electron spin, obtaining $\mathcal{F}=\mathcal{L}+s_e$. We then include the antiproton's spin to obtain the total angular momentum $\mathcal{J}=\mathcal{F}+s_{\overline{p}}$ \footnote{The choice of letters denoting sums of angular momenta is inconsistent in the literature, e.g., in Ref.~\cite{Bak98} $J=\mathcal{L}+s_e$ and $F$ is the total angular momentum, while in Ref.~\cite{Fri11} the authors also use the letter $G$. We follow the convention used in Refs.~\cite{Pas09,KorBak01}, using a curly font to avoid possible ambiguity.}. This addition scheme introduces the hyperfine structure shown in Fig.\ \ref{fig:hfs}. We may characterise any hyperfine state in the $(37,35)$ manifold using three numbers: $\mathcal{J}$, $\widetilde{m}_{\mathcal{J}}$, and $\mathcal{F}$. We build these states using the Clebsch-Gordan coefficients $C^{J, m_J}_{j_1, m_1; j_2, m_2}$:
\begin{align} 
|\mathcal{J},\widetilde{m}_{\mathcal{J}}; \mathcal{F}\rangle=&\sum_{\widetilde{m}_{\mathcal{F}}, \widetilde{m}_{s_{\overline{p}}}} C^{\mathcal{J},\widetilde{m}_{\mathcal{J}}}_{\mathcal{F},\widetilde{m}_{\mathcal{F}};s_{\overline{p}},\widetilde{m}_{s_{\overline{p}}}}|\mathcal{F},\widetilde{m}_{\mathcal{F}}\rangle |s_{\overline{p}},\widetilde{m}_{s_{\overline{p}}}\rangle\nonumber\\
=&\sum_{\widetilde{m}_{\mathcal{F}}, \widetilde{m}_{s_{\overline{p}}}}\sum_{\widetilde{m}_{\mathcal{L}}, \widetilde{m}_{s_{e}}} C^{\mathcal{J},\widetilde{m}_{\mathcal{J}}}_{\mathcal{F},\widetilde{m}_{\mathcal{F}};s_{\overline{p}},\widetilde{m}_{s_{\overline{p}}}}C^{\mathcal{F},\widetilde{m}_{\mathcal{F}}}_{L,\widetilde{m}_{\mathcal{L}};s_e,\widetilde{m}_{s_e}}\nonumber\\
&\times|35,\widetilde{m}_{\mathcal{L}}\rangle |s_e,\widetilde{m}_{s_e}\rangle |s_{\overline{p}},\widetilde{m}_{s_{\overline{p}}}\rangle.
\label{eqn:sol3}
\end{align}
Due to rotational invariance of the Hamiltonian and the exotic spin-dependent potentials considered below, the respective matrix elements do not depend on the specific $\widetilde{m}_{\mathcal{J}}$ value, so we denote the hyperfine structure states by $|\mathcal{J}; \mathcal{F}\rangle$.

Reference \cite{Pas09} presented the results of measurements of energies for the transitions $|35; 35.5\rangle \leftrightarrow |34;34.5\rangle$ (denoted as $\nu_{\text{HFS}}^{-}$) and $|36; 35.5\rangle \leftrightarrow |35; 34.5\rangle$ (denoted as $\nu_{\text{HFS}}^{+}$) along with the theoretically predicted values calculated in Refs.~\cite{KorBak01}. We compare them in Table \ref{tab:hfs} and present the values of $\Delta E$ -- a quantity constraining exotic interactions at 90\% acceptance level. We define it in such a way that
\begin{align} 
\int_{-\Delta E}^{+\Delta E} \frac{1}{\sqrt{2\pi} \sigma} e^{-(x-\mu)^2/(2\sigma^2)} dx=0.9,
\label{eqn:DeltaE}
\end{align}
where $\mu$ is the mean difference between theoretical and experimental transition energies, and $\sigma^2=\sigma_{th}^2+\sigma_{exp}^2$ ($\sigma_{th}$ and $\sigma_{exp}$ are here theoretical and experimental uncertainties, respectively). These values of $\Delta E$, characterising the level of agreement between theory and experiment taking into account the uncertainties of both, are used to constrain the exotic interactions.

\begin{table*}
\caption{\label{tab:hfs}Experimental and theoretical transition energies between hyperfine-structure states in the $(n,l)=(37,35)$ manifold, along with their differences and values of $\Delta E$, a parameter describing the level of agreement between theoretical and experimental results and taking into account their uncertainties. We define $\Delta E$ at the $90\%$ Confidence Level (C.L.) in Eq.\ (\ref{eqn:DeltaE}).}
\begin{ruledtabular}
\begin{tabular}{ccccc}
 & Experiment \cite{Pas09} & Theory \cite{KorBak01} & Difference & $\Delta E$ (at $90\%$ C.L.)\\ \hline
 $\nu_{\text{HFS}}^{+}$ & 12.896 641(63) GHz & 12.8963(13)  GHz & 0.3(1.3) MHz & 2.2 MHz\\
 $\nu_{\text{HFS}}^{-}$ & 12.924 461(63) GHz & 12.9242(13) GHz & 0.3(1.3) MHz & 2.2 MHz\\
\end{tabular}
\end{ruledtabular}
\end{table*}

\section{Spin-dependent potentials}
In Ref. \cite{Dob06}, Dobrescu and Mocioiu introduced 16 independent spin-spin interactions. This list is reviewed and corrected in \cite{FadIP}. For studies of exotic spin couplings using $^4$He$^{+} \overline{\text{p}}$, only those interactions that are invariant under spatial inversion and time reversal are relevant. These two conditions allow shifts of energy levels in first-order perturbation theory. There are five spin-dependent potentials that satisfy these requirements: two velocity-independent potentials and three velocity-dependent potentials. In the coordinate-space representation they have the form \cite{FadIP}

\begin{widetext}
\begin{eqnarray}
		V_2&=&f_2^{e\overline{p}}  \frac{\hbar c}{\pi} \left(\textbf{s}_{\overline{p}}\cdot\textbf{s}_{e}\right) \frac{e^{-r/\lambdabar}}{r},\label{eq:v2}\\
		V_3&=&f_3^{e\overline{p}} \frac{\hbar^3}{\pi m_e^2 c} \left[\textbf{s}_{\overline{p}}\cdot\textbf{s}_e\left(\frac{1}{\lambdabar r^2}+\frac{1}{r^3}+\frac{4\pi}{3}\delta^3(r)\right)-\left(\textbf{s}_{\overline{p}}\cdot\textbf{r}\right)\left(\textbf{s}_e\cdot\textbf{r}\right)\left(\frac{1}{\lambdabar^2 r^3}+\frac{3}{\lambdabar r^4}+\frac{3}{r^5}\right)\right]e^{-r/\lambdabar},\label{eq:v3}\\
		V_{4-5}&=&f_{4-5}^{e\overline{p}}  \frac{i \hbar^3}{4 m_e^2 c} \textbf{s}_{\overline{p}}\cdot \left[\left(\frac{m_e}{m_{\overline{p}}+m_e}\nabla_{\overline{p}}-\frac{m_{\overline{p}}}{m_{\overline{p}}+m_e}\nabla_e\right)\times\textbf{r},\left(\frac{1}{r^3}+\frac{1}{\lambdabar r^2}\right)e^{-r/\lambdabar}\right]_{+},\label{eq:v4}\\
		V_{4+5}&=&f_{4+5}^{e\overline{p}} \frac{i \hbar^3}{4 m_e^2 c}  \textbf{s}_e \cdot \left[\left(\frac{m_e}{m_{\overline{p}}+m_e}\nabla_{\overline{p}}-\frac{m_{\overline{p}}}{m_{\overline{p}}+m_e}\nabla_e\right)\times\textbf{r},\left(\frac{1}{r^3}+\frac{1}{\lambdabar r^2}\right)e^{-r/\lambdabar}\right]_{+},\label{eq:v5}\\
		V_8&=&-f_8^{e\overline{p}} \frac{\hbar^3}{4 \pi m_e^2 c} \left[\textbf{s}_{e}\cdot\left(\frac{m_e}{m_{\overline{p}}+m_e}\nabla_{\overline{p}}-\frac{m_{\overline{p}}}{m_{\overline{p}}+m_e}\nabla_e\right),\left[\textbf{s}_{\overline{p}}\cdot\left(\frac{m_e}{m_{\overline{p}}+m_e}\nabla_{\overline{p}}-\frac{m_{\overline{p}}}{m_{\overline{p}}+m_e}\nabla_e\right), \frac{e^{-r/\lambdabar}}{r}\right]_{+}\right]_{+},\label{eq:v8}\label{eq:v8}
\end{eqnarray}
\end{widetext}
where $f_i^{e\overline{p}}$ is the dimensionless coupling parameter of the $i$-th interaction between the electron and the antiproton, $\textbf{r}=\textbf{r}_e-\textbf{r}_{\overline{p}}$ is the position vector directed from the antiproton to the electron, $r$ is the distance between the electron and antiproton, $\nabla_{\overline{p}}$ and $\nabla_e$ are vector differential operators in coordinate space of the antiproton and the electron, respectively, and  $\textbf{s}_{\overline{p}}$ and $\textbf{s}_e$ are the spins of the antiproton and the electron, respectively. By $[\cdot,\cdot]_{+}$ we denote an anticommutator. 

The potentials $V_{4-5}$ and $V_{4+5}$ have exactly the same orbital part and differ only in the spin part (they contain antiprotonic and electronic spin, respectively). We are interested in the states with high orbital number ($\mathcal{L}=35$) and total angular momentum ${\cal J} \approx \mathcal{L}$. For such states both spins are either (almost) parallel, or antiparallel to ${\bf \mathcal{L}}$. We are considering the transitions $\nu_\mathrm{HFS}^{\pm}$ (see Fig.\ \ref{fig:hfs}), where spins may flip, but the orbital part does not change. Thus, we can say that each of the potentials $V_{4\pm 5}$ contributes only to the transition where the respective spin flips. To see how the spins ${\bf s}_{\bar p}$ and ${\bf s}_{e}$ behave in the transitions $\nu_\mathrm{HFS}^{\pm}$, we need to expand the four states in question according to Eq.\ (\ref{eqn:sol3}). The Clebsch-Gordon coefficients take the simplest form for the states with maximum projection $\tilde m_{\cal I}$. The amplitudes of the states with various spin projections are presented in Table \ref{tab:amplitudes}. The transition $\nu_\mathrm{HFS}^{+}$ links the first pair of states, while $\nu_\mathrm{HFS}^{-}$ links the last pair. In the first approximation, both transitions correspond to an electron spin flip --- the admixture of the spin flip of the antiproton is suppressed roughly by two orders of magnitude. We see that the expectation value of the potential $V_{4-5}$ is practically the same for the upper and lower states and, therefore, the transition frequencies $\nu_\mathrm{HFS}^{\pm}$ are practically not affected by this potential, so we are not constraining it. For all other potentials, the electron spin flip causes the sign-change of their expectation values. In the following, we focus on the potentials $V_2$, $V_3$, $V_{4+5}$, and $V_8$.

\begin{table}
\caption{\label{tab:amplitudes} Amplitudes of the states with different spin projections. The first arrow corresponds to the projection of the antiproton spin and the second one denotes the projection of the electron spin.}
\begin{ruledtabular}
\begin{tabular}{c|cccc}
 $|{\cal J},\tilde m_{\cal J}; {\cal F}\rangle$ & $(\uparrow,\uparrow)$ & $(\uparrow,\downarrow)$ & $(\downarrow,\uparrow)$ & $(\downarrow,\downarrow)$ \\ \hline
 $|35,35;34\tfrac12\rangle$ &$-0.1187$ &$ 0.9929$&$   0   $ &$   0   $ \\
  $|36,36;35\tfrac12\rangle$ &$   1   $ &$   0   $&$    0  $ &$    0  $ \\ \hline
 $|34,34;34\tfrac12\rangle$ &$ 0.0201$ &$-0.1178$&$-0.1178$ &$ 0.9858$ \\
 $|35,35;35\tfrac12\rangle$ &$-0.1170$ &$-0.0140$&$ 0.9930$ &$    0  $
\end{tabular}
\end{ruledtabular}
\end{table}


\section{Results}\label{sec:results}
\begin{figure*}
\subfloat[\label{fig:v2}]{\includegraphics[width=.45\linewidth]{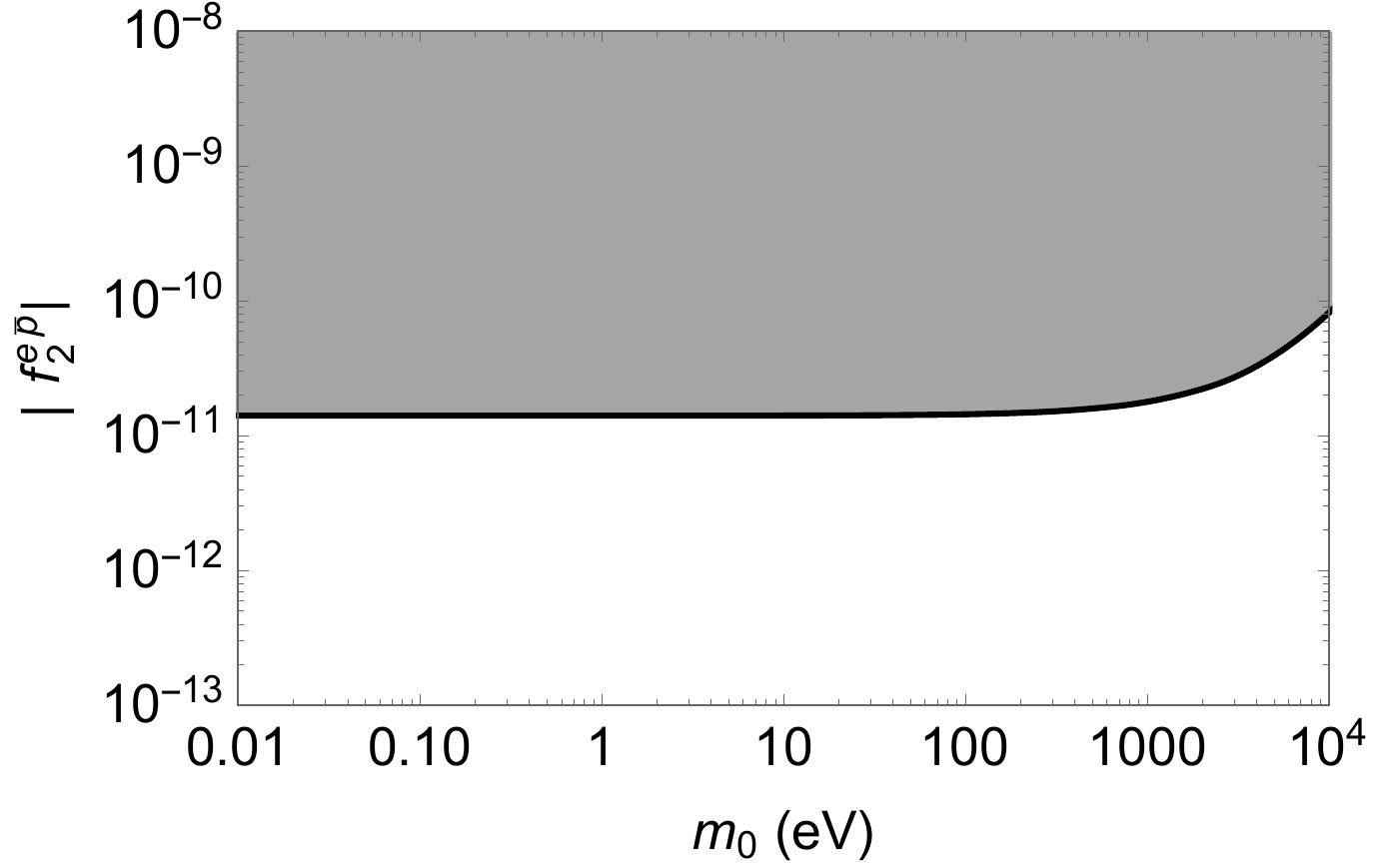}}
\subfloat[\label{fig:v3}]{\includegraphics[width=.45\linewidth]{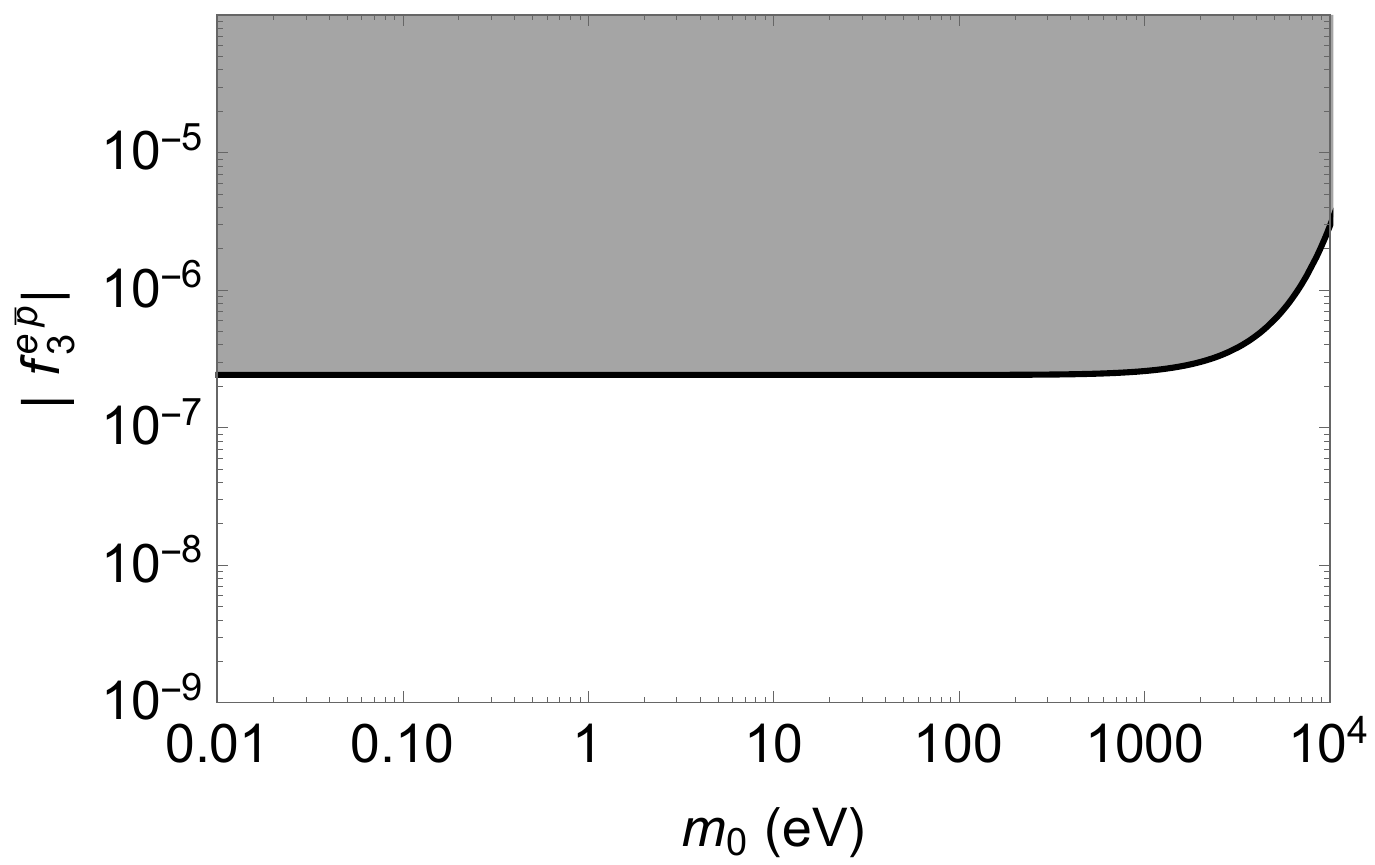}}\hfill
\subfloat[\label{fig:v4p5}]{\includegraphics[width=.45\linewidth]{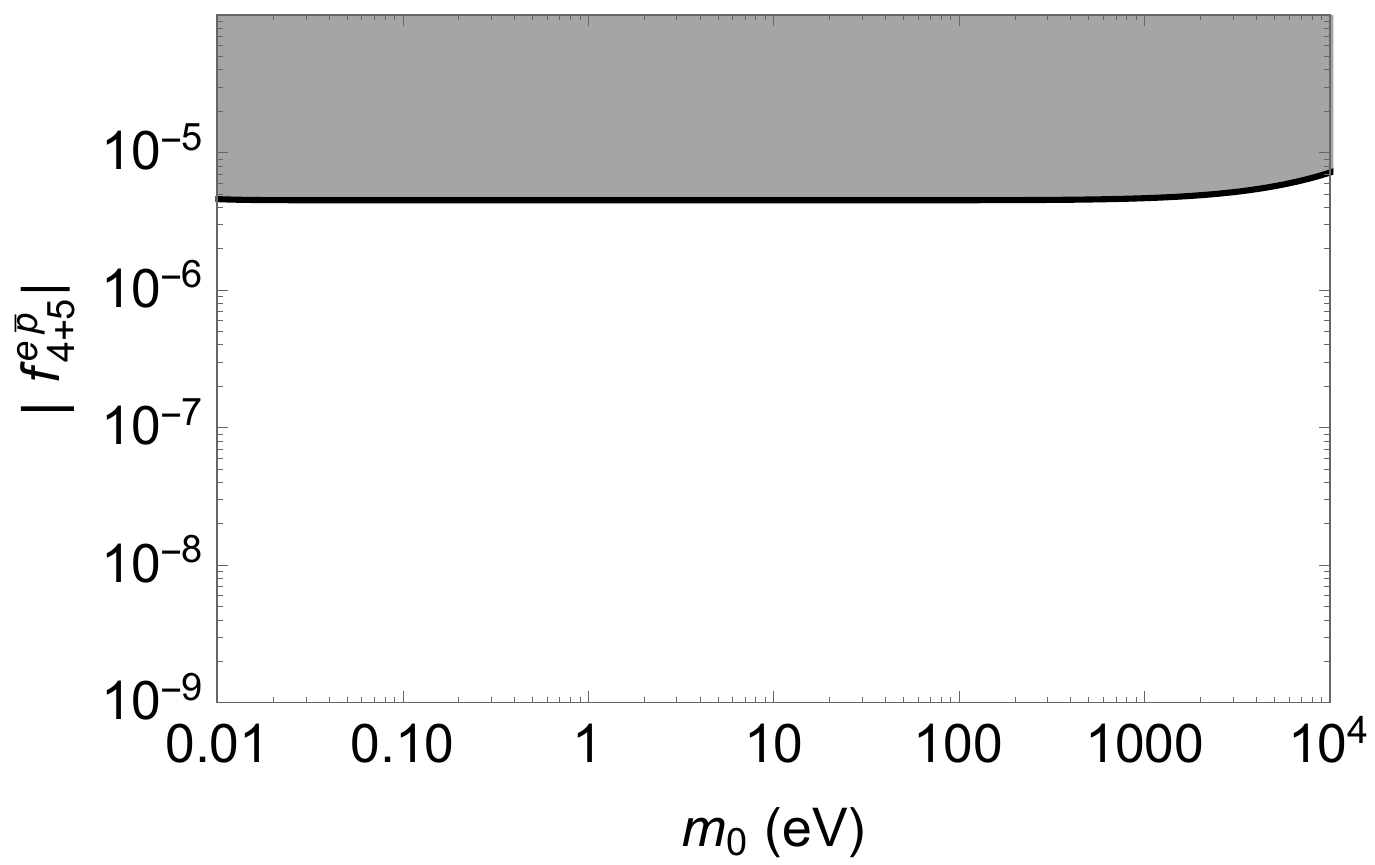}}
\subfloat[\label{fig:v8}]{\includegraphics[width=.45\linewidth]{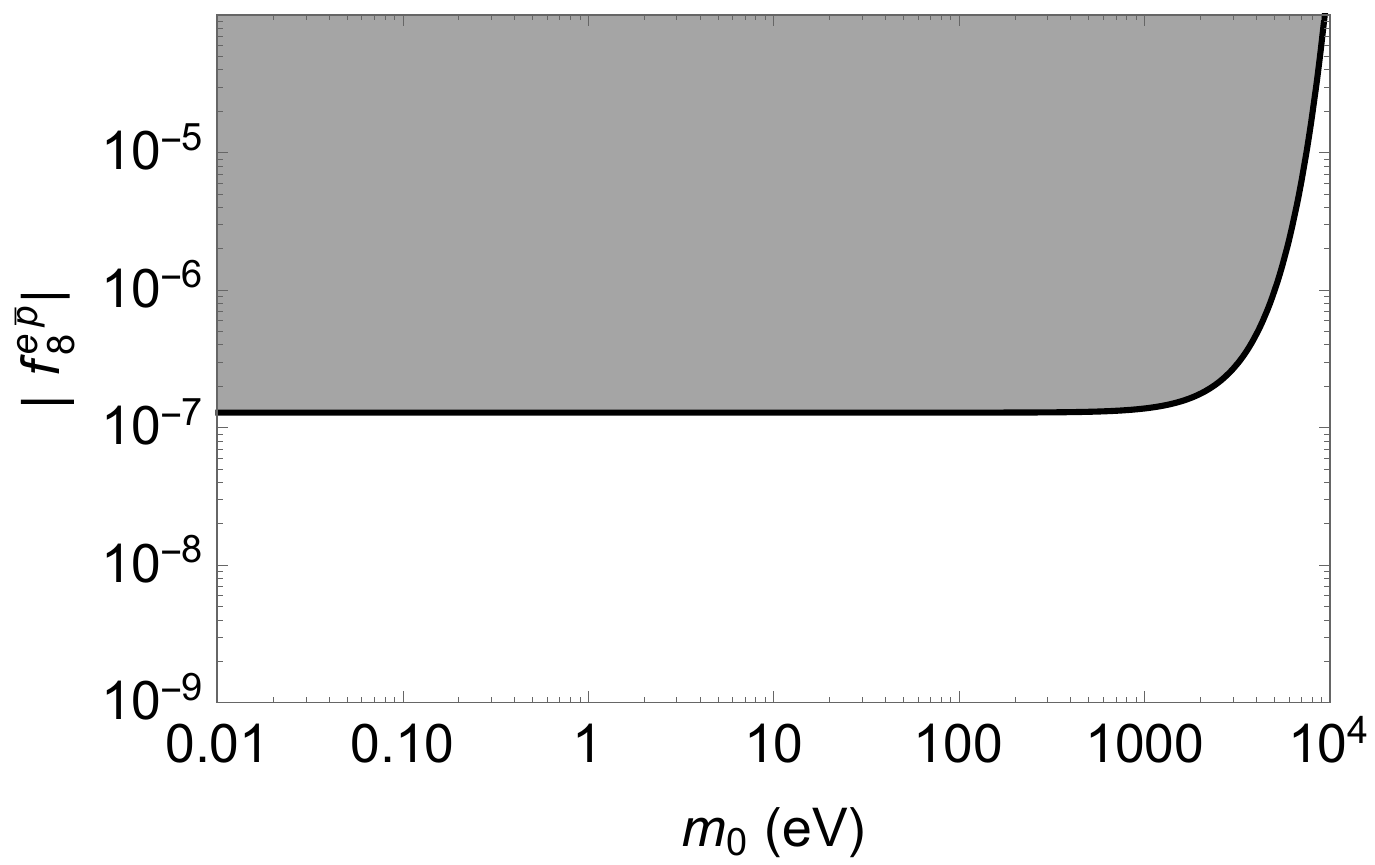}}\hfill
\caption{Constraints (at the 90$\%$ confidence level) on the magnitude of the dimensionless coupling constants $f_i^{e\overline{p}}$ as a function of the boson mass $m_0$.}
\label{fig:results}
\end{figure*}

For every considered potential $V_i$, we introduce the operator $\mathcal{V}_i$, defined as $V_i=f_i^{e\overline{p}} \mathcal{V}_i$. Then we may estimate the energy shift between states $|\mathcal{J}_a; \mathcal{F}_a\rangle$ and $|\mathcal{J}_b; \mathcal{F}_b\rangle$ caused by a $\mathcal{V}_i$ operator using first-order perturbation theory and the approximate wavefunctions as follows:
\begin{align}
\Delta U_{ab,i}(m_0)=&\langle \mathcal{J}_a;\mathcal{F}_a| \mathcal{V}_i(m_0) |\mathcal{J}_a;\mathcal{F}_a\rangle \nonumber\\
&- \langle \mathcal{J}_b;\mathcal{F}_b| \mathcal{V}_i(m_0) |\mathcal{J}_b;\mathcal{F}_b\rangle,
\label{eq:Uabi}
\end{align}
where $\mathcal{V}_i$ depends on the intermediate boson mass $m_0$, as can be seen in Eqs. (\ref{eq:v2}) -- (\ref{eq:v8}). For given values of the $f_i^{e\overline{p}}$ parameter and boson mass $m_0$, the exotic potential causes a shift of the transition energy equal to $f_i^{e\overline{p}} \Delta U_{ab,i}(m_0)$. The maximal discrepancy between theory and experiment is equal to $\Delta E$ (see Table \ref{tab:hfs}), so for any value of $m_0$, the inequality
\begin{equation}
\left| f_i^{e\overline{p}}(m_0) \Delta U_{ab,i}(m_0) \right| \leq \Delta E
\label{eq:DeltaE}
\end{equation}
holds. The constraints on the $f_i^{e\overline{p}}$ parameter values can be calculated as
\begin{equation}
|f_i^{e\overline{p}}(m_0)|\leq\left|\frac{\Delta E}{\Delta U_{ab,i}(m_0)}\right|.
\label{eq:gege}
\end{equation}
To obtain constraints on $f_{i}^{e\overline{p}}$ as a function of $m_0$, we perform numerical calculations of $\Delta U_{ab,i}$ for several $m_0$ values and then interpolate between them to obtain continuous exclusion plots.  We perform this procedure for both transitions, $\nu_{\rm HFS}^+$ and $\nu_{\rm HFS}^-$, and choose the more stringent of the two constraints. The obtained constraints are presented in Fig. \ref{fig:results}. 

As can be seen in the exclusion plots, for bosons with masses larger than several keV/$c^2$, the constraints weaken. This is explained by the fact that our system is less sensitive to interactions mediated by bosons with a Compton wavelength much shorter than the size of the antiprotonic helium atom.

We test our numerically derived constraints by comparing them with results of theoretical estimates (Table \ref{tab:results}). For these considerations, we use atomic units ($\hbar=m_e=|e|=1$) and explore the limit of zero boson mass ($\lambdabar\to\infty$). The fact that the speed of light is present in potential $V_2$ as $c$ (in atomic units $c = 1/\alpha \approx 137$), while in the rest of the potentials it comes as $c^{-1}$ suggests that the constraints on $V_2$ should be approximately $\alpha^{-2} \sim 10^4$ times more stringent than on the other potentials. Additionally, one may show (see Appendix \ref{sec:v45}), that due to the spherical symmetry of the electron wavefunction, for the considered system in potentials $V_{4\pm 5}$ only the terms containing derivatives over the antiproton position are relevant. These terms are suppressed by the factor $m_e/(m_{\overline{p}}+m_e)\approx0.5\times10^{-3}$. Using the virial theorem and taking the potential energy to be $\sim$ 1 a.u.\ for an antiproton with $n\approx 35$, we may estimate $\langle \nabla_{\overline{p}}\rangle \sim \sqrt{m_{\overline{p}}}$, which yields
\begin{align}
\left\langle \frac{m_e}{m_{\overline{p}}+m_e}\nabla_{\overline{p}}\right\rangle \sim \frac{m_e}{\sqrt{m_{\overline{p}}}} \approx 0.02.
\label{eq:nablap}
\end{align}
The other quantities present in the potentials, such as the spins, $\textbf{s}_{\overline{p}}$ and $\textbf{s}_e$, particle positions, $\textbf{r}_{\overline{p}}$ and $\textbf{r}_e$, and the differential operator $\nabla_e$ can be considered to be of order unity. Comparing the approximate expectation values of the potentials with the value of $\Delta E\lesssim 2~{\rm MHz} \approx 3 \times 10^{-10}~{\rm a.u.}$ from Table \ref{tab:hfs} yields the approximate constraints presented in Table \ref{tab:results}. These constraints are similar to the ones coming from numerical integration.
\begin{table}
\caption{\label{tab:results}Numerical calculations and order-of-magnitude estimates for constraints on the $|f_i^{e\overline{p}}|$ parameters in the massless boson limit.}
\begin{ruledtabular}
\begin{tabular}{c|cccc}
& $|f_2|$ & $|f_3|$ & $|f_{4+5}|$ & $|f_8|$ \\ \hline
Estimates & $3\times 10^{-12}$ & $3\times 10^{-8}$ & $2\times 10^{-6}$ & $3\times 10^{-8}$ \\
Numerics & $1.4\times 10^{-11}$ & $2.5\times 10^{-7}$& $4.4\times 10^{-6}$ & $1.3\times 10^{-7}$
\end{tabular}
\end{ruledtabular}
\end{table}



\section{Summary and Outlook}
For the first time, semileptonic spin-dependent interactions between matter and antimatter have been constrained. We did so by investigating hypothetical antiproton-electron spin-dependent interactions in antiprotonic helium. Moreover, this analysis provides the first constraints on velocity-dependent spin-dependent matter-antimatter interactions. Our constraints were obtained by comparing theoretical predictions and laboratory results, together with our calculated expectation values of exotic potentials. The current accuracy of the experiment \cite{Pas09} is 20 times higher than the accuracy of the theory \cite{KorBak01}. Further improvement in the theory can improve limits obtained in the present work by an order of magnitude. 

\begin{acknowledgments}
The authors acknowledge Masaki Hori and Victor Korobov for generously sharing their knowledge about antiprotonic helium experiments and calculations, and Szymon Pustelny for his insights and opinions on the manuscript. F.F. would like to thank Konrad Szyma\'{n}ski and Roman Skibi\'{n}ski for their useful remarks and ideas. M.K. is grateful to the Mainz Institute for Theoretical Physics (MITP) for its hospitality.
This project was partially supported by the Polish Ministry of Science and Higher Education within the Diamond Grant (Grant No. 0143/DIA/2016/45), the U.S. National Science Foundation under grant PHY-1707875, Russian Foundation for Basic Research under Grant No. 17-02-00216, Humboldt Research Fellowship, and a mini-grant from FQXi the  Foundational Questions Institute.
\end{acknowledgments}

\appendix

\section{Approximate antiprotonic helium spatial wavefunctions}\label{sec:wave}
\begin{figure}
\includegraphics[width=0.4\textwidth]{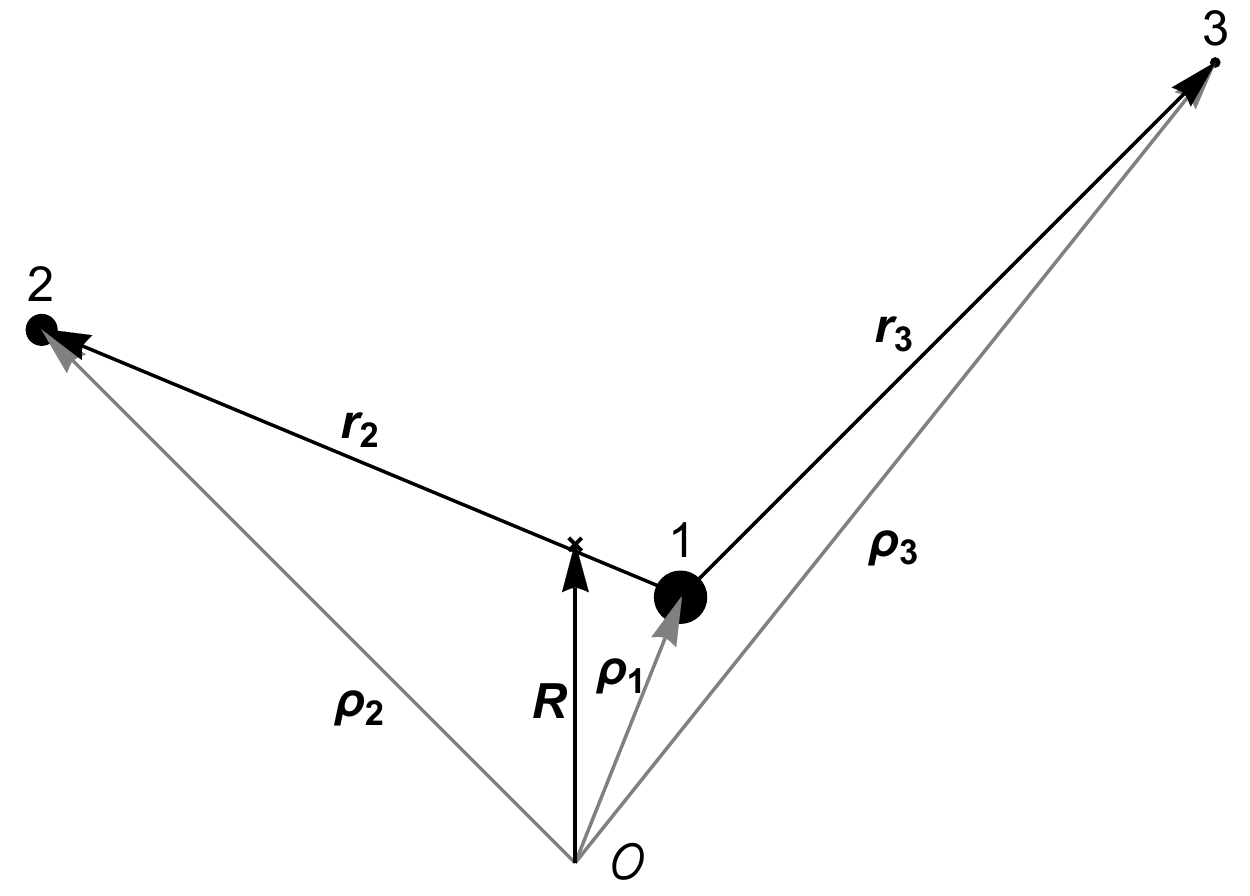}\\
\caption{\label{fig:3body}Schematic diagram of the three-body system described in this section.}
\end{figure}

Consider a system of three particles with masses $m_1$, $m_2$, and $m_3$ and respective charges $q_1$, $q_2$, and $q_3$, as shown in Fig.\ \ref{fig:3body}. The positions of these particles with respect to an arbitrary point $O$ is denoted by  $\boldsymbol{\rho}_1$, $\boldsymbol{\rho}_2$, and $\boldsymbol{\rho}_3$, respectively. If these particles interact only electrostatically, the energy of this system is
\begin{align} 
E=&\frac{1}{2}m_1 \dot{\boldsymbol{\rho}}_1^2+\frac{1}{2}m_2 \dot{\boldsymbol{\rho}}_2^2+\frac{1}{2}m_3 \dot{\boldsymbol{\rho}}_3^2+\frac{q_1 q_2}{|\boldsymbol{\rho}_1-\boldsymbol{\rho}_2|}\nonumber\\
&+\frac{q_1 q_3}{|\boldsymbol{\rho}_1-\boldsymbol{\rho}_3|}+\frac{q_2 q_3}{|\boldsymbol{\rho}_2-\boldsymbol{\rho}_3|}.
\label{eqn:energy1}
\end{align}
Let us introduce a new coordinate system with the positions defined as
\begin{align} 
\mathbf{R}=&\frac{m_1 \boldsymbol{\rho}_1+m_2 \boldsymbol{\rho}_2+m_3 \boldsymbol{\rho}_3}{M},\\
\mathbf{r_2}=&\boldsymbol{\rho}_2-\boldsymbol{\rho}_1,\\
\mathbf{r_3}=&\boldsymbol{\rho}_3-\boldsymbol{\rho}_1,
\label{coords}
\end{align}
where the total mass is $M=m_1+m_2+m_3$. In these coordinates, the energy of the system is
\begin{align} 
E=&\frac{1}{2}M \dot{\mathbf{R}}^2-\frac{m_2 m_3}{M} \dot{\mathbf{r}}_2\cdot \dot{\mathbf{r}}_3+\frac{m_2(m_1+m_3)}{2M} \dot{\mathbf{r}}_2^2\nonumber\\
&+\frac{m_3(m_1+m_2)}{2M}\dot{\mathbf{r}}_3^2+\frac{q_1 q_2}{|\mathbf{r}_2|}+\frac{q_1 q_3}{|\mathbf{r}_3|}+\frac{q_2 q_3}{|\mathbf{r}_2-\mathbf{r}_3|}.
\label{eqn:energy2}
\end{align}

Let us now apply these general considerations to the antiprotonic helium atom. Let particle 1 be the helium nucleus ($\alpha$ particle), particle 2 be the antiproton ($\overline{p}$) and particle 3 be the electron ($e$). Then we have $m_\textrm{nucl},m_{\overline{p}} \gg m_e$ and $M \approx m_\textrm{nucl}+m_{\overline{p}}$. The approximate energy is:
\begin{align} 
E\approx&\frac{1}{2}M \dot{\mathbf{R}}^2+\frac{m_\textrm{nucl} m_{\overline{p}}}{2M} \dot{\mathbf{r}}_{\overline{p}}^2+\frac{m_e}{2}\dot{\mathbf{r}}_e^2-\frac{2e^2}{|\mathbf{r}_{\overline{p}}|}-\frac{2e^2}{|\mathbf{r}_e|}\nonumber\\
&+\frac{e^2}{|\mathbf{r}_{\overline{p}}-\mathbf{r}_e |}.
\label{eqn:energy3}
\end{align}
The first term may be eliminated by going to the center-of-mass frame. We re-write the remaining terms as the Hamiltonian:
\begin{align} 
H=&\frac{1}{2\mu_{\overline{p}}} \mathbf{p}_{\overline{p}}^2+\frac{1}{2 m_e}\mathbf{p}_e^2-\frac{2e^2}{|\mathbf{r}_{\overline{p}}|}-\frac{2e^2}{|\mathbf{r}_e|}+\frac{e^2}{|\mathbf{r}_{\overline{p}}-\mathbf{r}_e |},
\label{eqn:energy4}
\end{align}
where $\mu_{\overline{p}}=m_{\textrm{nucl}} m_{\overline{p}}/(m_\textrm{nucl}+m_{\overline{p}})$ is the reduced mass of the antiproton and $\mathbf{p}$ are the momenta of the relevant particles. After quantization, we obtain
\begin{align} 
\hat{H}=\left(-\frac{\hbar^2}{2\mu_{\overline{p}}} \nabla_{\overline{p}}^2-\frac{2e^2}{|\mathbf{r}_{\overline{p}}|}\right)+\left(-\frac{\hbar^2}{2 m_e}\nabla_e^2-\frac{2e^2}{|\mathbf{r}_e|}\right)+\frac{e^2}{|\mathbf{r}_{\overline{p}}-\mathbf{r}_e |}.
\label{eqn:hamiltonian}
\end{align}

Let us neglect the last term for a moment. Then the problem described by such a Hamiltonian can be solved precisely as two decoupled hydrogen-like atoms. The solution is characterised by two sets of three quantum numbers (principal quantum number $n$, orbital angular momentum quantum number $l$, and magnetic quantum number $\widetilde{m}$), one set for the antiproton and one set for the electron. We may write the wavefunction as
\begin{align} 
\Psi=\psi^{(\overline{p})}_{n_{\overline{p}},l_{\overline{p}},\widetilde{m}_{\overline{p}}} \psi^{(e)}_{n_e,l_e,\widetilde{m}_e},
\label{eqn:solution1}
\end{align} 
where $\psi^{(a)}_{n,l,\widetilde{m}}$ is a generalised hydrogen-atom wavefunction \cite{Gri} for particle $a$:
\begin{align} 
\psi^{(a)}_{n,l,m}&(r,\theta,\phi)=\sqrt{\frac{4 (Z^{(a)})^3\mu^3 (n-l-1)!}{n^4 (n+l)!}} \left(\frac{2Z^{(a)} \mu^{(a)} r}{n}\right)^l \nonumber\\
&\times e^{-\frac{Z^{(a)} \mu^{(a)} r}{n}}L^{2l+1}_{n-l-1} \left(\frac{2Z^{(a)} \mu^{(a)} r}{n}\right) Y_{l}^{m}(\theta,\phi).
\label{eqn:solution2}
\end{align} 
In this formula, $\mu^{(a)}$ denotes the reduced mass of particle $a$ (in our case, $\mu^{(e)}\approx m_e$), $Z^{(a)}$ is the effective charge seen by particle $a$ [for Hamiltonian (\ref{eqn:hamiltonian}) without the last term, we would have $Z^{(\overline{p})}=Z^{(e)}=2$], $L^{2l+1}_{n-l-1}$ is the generalised Laguerre polynomial, and $Y_{l}^{m}$ is the spherical harmonic function. We will focus on states where the electron is in the ground state $(1,0,0)$, while the antiproton can be in an arbitrary state $(n_{\overline{p}},l_{\overline{p}},\widetilde{m}_{\overline{p}})$. Let us define a state
\begin{align} 
| \Psi_{n_{\overline{p}},l_{\overline{p}},\widetilde{m}_{\overline{p}}}(Z^{(\overline{p})},Z^{(e)}) \rangle=\psi^{(\overline{p})}_{n_{\overline{p}},l_{\overline{p}},\widetilde{m}_{\overline{p}}} \psi^{(e)}_{1,0,0},
\label{eqn:solution3}
\end{align} 
where the coefficients $Z^{(\overline{p})}$ and $Z^{(e)}$ are effective charges in $\psi^{(\overline{p})}_{n_{\overline{p}},l_{\overline{p}},\widetilde{m}_{\overline{p}}}$ and $\psi^{(e)}_{1,0,0}$, respectively.

In order to account for the effects of the electromagnetic interaction between the antiproton and electron, represented by the last term in Eq.\ (\ref{eqn:hamiltonian}), we use the variational method. The other orbiting particle screens the nuclear charge, so we treat the charges $Z^{(\overline{p})}$ and $Z^{(e)}$ as variational parameters. Let us point out that even if we restrict ourselves to states with $l_{\overline{p}}=35$, the variational method shall give us an approximation of the state with a minimal energy having this $l_{\overline{p}}$ value, i.e., the state $(36,35)$ \cite{Bethe}. To get an approximation for the state $(37,35)$, we shall use a test function orthogonal to the found approximation of the $(36,35)$ state \cite{Sak}.

Using the variational method, we find the values of $Z^{(\overline{p})}_{36}$ and $Z^{(e)}_{36}$ that minimize the energy of the system $\langle \Psi_{36,35,35}(Z^{(\overline{p})}_{36},Z^{(e)}_{36})| \hat{H} | \Psi_{36,35,35}(Z^{(\overline{p})}_{36},Z^{(e)}_{36})\rangle$ (in fact, the specific value of $\widetilde{m}_{\overline{p}}$ does not matter, as long as $|\widetilde{m}_{\overline{p}}|\leq 35$; we choose it to be 35 to simplify the calculations). In order to estimate the uncertainty of our wavefunctions, the energy obtained this way is compared with the one obtained with a more accurate method (Table \ref{tab:energy}). 

Let us define $|\Phi_{36}\rangle := | \Psi_{36,35,35}(Z^{(\overline{p})}_{36},Z^{(e)}_{36})\rangle$. Now we construct a test function with which we approximate the state $(37,35)$:
\begin{align} 
|\widetilde{\Phi}_{37}(Z^{(\overline{p})},Z^{(e)}) \rangle=&|\Psi_{37,35,35}(Z^{(\overline{p})},Z^{(e)}) \rangle\nonumber\\
&-\langle\Phi_{36} | \Psi_{37,35,35}(Z^{(\overline{p})},Z^{(e)}) \rangle |\Phi_{36}\rangle.
\label{eqn:testfunction}
\end{align}
We find the parameters $Z^{(\overline{p})}_{37}$ and $Z^{(e)}_{37}$ that minimise the energy $\langle \widetilde{\Phi}_{37}(Z^{(\overline{p})}_{37},Z^{(e)}_{37})| \hat{H} | \widetilde{\Phi}_{37}(Z^{(\overline{p})}_{37},Z^{(e)}_{37})\rangle$ (we again compare it with the more precise result in Table \ref{tab:energy}). By substituting the values of these parameters into Eq.\ (\ref{eqn:testfunction}), we finally obtain the approximate wavefunction for the $(37,35)$ state as:
\begin{align} 
|&\Psi_{37,35,\widetilde{m}_{(\overline{p})}} \rangle=|\Psi_{37,35,\widetilde{m}_{(\overline{p})}}(Z^{(\overline{p})}_{37},Z^{(e)}_{37})\rangle\nonumber\\
&-\langle\Phi_{36} | \Psi_{37,35,35}(Z^{(\overline{p})}_{37},Z^{(e)}_{37}) \rangle |\Psi_{36,35,\widetilde{m}_{(\overline{p})}}(Z^{(\overline{p})}_{36},Z^{(e)}_{36})\rangle.
\label{eqn:wavefunction}
\end{align}
The value of $\langle\Phi_{36} | \Psi_{37,35,35}(Z^{(\overline{p})}_{37},Z^{(e)}_{37}) \rangle$ is the $\beta$ constant in Eq.\ (\ref{eqn:sol1}). We find that $\beta\approx0.216$.

\begin{table}
\caption{\label{tab:energy}Values of the ionization energies calculated with approximate wavefunctions for distinct antiprotonic helium states (in hartrees).}
\begin{ruledtabular}
\begin{tabular}{ccc}
$(n,l)$ & This paper & Ref. \cite{Yam02} \\ \hline
 $(36,35)$ & -2.979 & -2.984  \\
 $(37,35)$ & -2.883 & -2.899 \\
\end{tabular}
\end{ruledtabular}
\end{table}

\section{Variational method sensitivity check}\label{sec:check}
In the main text, we use the antiprotonic helium wavefunctions derived in Appendix \ref{sec:wave}. As mentioned there, we use the variational method to find the wavefunction of the first excited state with angular momentum $\mathcal{L}=35$. We note that in general this method is used to find the ground state, so in this Appendix we will check whether or not our final results are sensitive to the precise values of the effective charges. 

In principle, we could do this by guessing any reasonable values of the effective charges $Z$, but instead we will use knowledge coming from Appendix \ref{sec:wave}, i.e., the approximate spatial wavefunction (\ref{eqn:testfunction}) denoted here by $\Psi$ (which is also a function of the effective charges) and the values of the effective charges $Z_\textrm{var}$, to perform an educated guess.

The sensitivity check is based on the observation that the effective charge that acts on one of the particles decreases with distance continuously from the maximum value of 2, when this particle is much closer to the nucleus than the other particle, to the minimum value of 1, when this particle is much further away from the nucleus than the other particle. To model this screening behaviour, we use a function of the form
\begin{align} 
Z(r)=1+\frac{1}{1+\left(\frac{r}{b}\right)^2},
\label{eqn:Z}
\end{align}
that will give us the effective charge acting on a particle at a distance $r$ from the nucleus while the other particle is at the distance parametrised by $b$ (which may have different values for the antiproton and electron). We define the parameter $b_e$ for the electron (and analogously $b_{\overline{p}}$ for the antiproton), as the radius of a sphere centred at the nucleus, for which the probabilities of finding the antiproton (and analogously the electron) outside and inside the sphere are equal:
\begin{align} 
\int_{\mathbb{R}^3} \int_0^{b_e} \int_0^{\pi} \int_0^{2\pi} r_{\overline{p}}^2 \sin{\theta_{\overline{p}}} \Psi^\ast \Psi ~ d\mathbf{r}_e ~ dr_{\overline{p}} ~ d\theta_{\overline{p}} ~ d\phi_{\overline{p}}=\frac{1}{2}.
\label{eqn:b}
\end{align}

\begin{figure}
\includegraphics[width=0.4\textwidth]{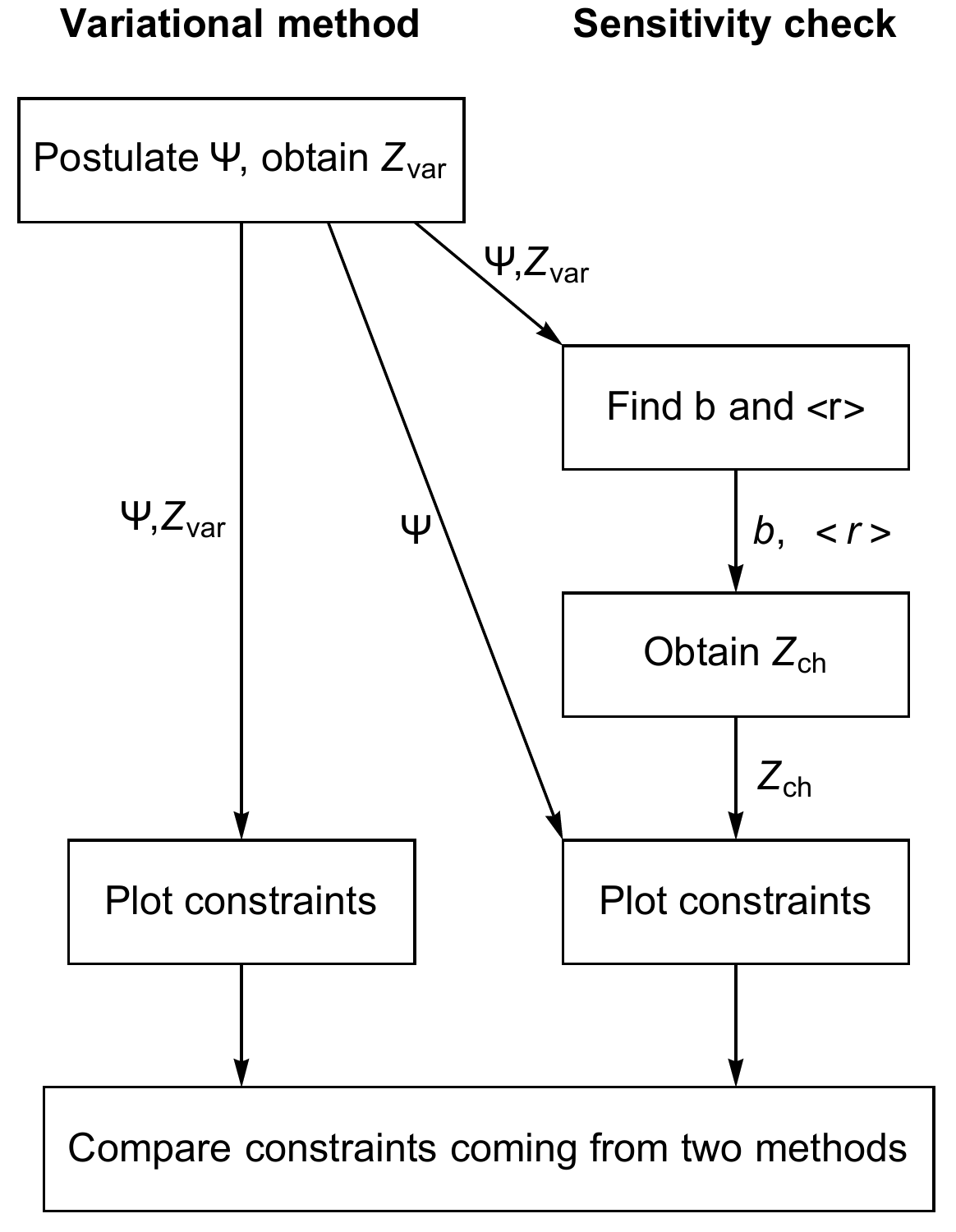}\\
\caption{\label{fig:diagram}Diagram illustrating the steps of the variational method described in Appendix \ref{sec:wave} and the sensitivity check procedure discussed in Appendix \ref{sec:check} to verify the accuracy of the calculations.}
\end{figure}

The plan of the calculations is as follows (see Fig. \ref{fig:diagram}). At first, we use the wavefunction $\Psi$ with the effective charges $Z_\textrm{var}$ to calculate the values of the parameters $b$ and the mean distances from the nucleus $\langle r \rangle$ for both of the particles. Then we estimate the new effective charges $Z_\textrm{ch}=Z(\langle r \rangle)$. In the end, we use the wavefunction $\Psi$ with the effective charges $Z_\textrm{ch}$ to obtain new constraints on the exotic potentials. Eventually, if the constraints obtained with $Z_\textrm{ch}$ match (to the desired level of accuracy) the constraints obtained with $Z_\textrm{var}$, we may conclude that our results are not significantly altered by the use of the variational method.

The value of $\langle r \rangle$ is calculated via the expression
\begin{align} 
\langle r_e\rangle = \int_{\mathbb{R}^3} \int_{\mathbb{R}^3} r_e \Psi^\ast \Psi ~ d\mathbf{r}_e ~ d\mathbf{r}_{\overline{p}}
\label{eqn:meanr}
\end{align}
for the electron, and analogously for the antiproton. We also calculate the parameters $b_e$ and $b_{\overline{p}}$ using Eq.\ (\ref{eqn:b}). We input the values of $b$ and $\langle r \rangle$ obtained for both of the particles into Eq.\ (\ref{eqn:Z}) with $\langle r \rangle$ as an argument. This results in two numbers, which will be our new effective charges $Z_\textrm{ch}$. The values of these charges, together with the effective charges coming from the variational method are presented in Table \ref{tab:charges}.
\begin{table}
\caption{\label{tab:charges}Comparison of the effective charges for the $(n,l)=(37,35)$ state obtained with the variational method and sensitivity check.}
\begin{ruledtabular}
\begin{tabular}{ccc}
 & Variational method & Sensitivity check \\ \hline
$Z^{(e)}$ & 1.30 & 1.19  \\
 $Z^{(\overline{p})}$ & 1.80 & 1.78 \\
\end{tabular}
\end{ruledtabular}
\end{table}
 
\begin{figure}
\includegraphics[width=0.4\textwidth]{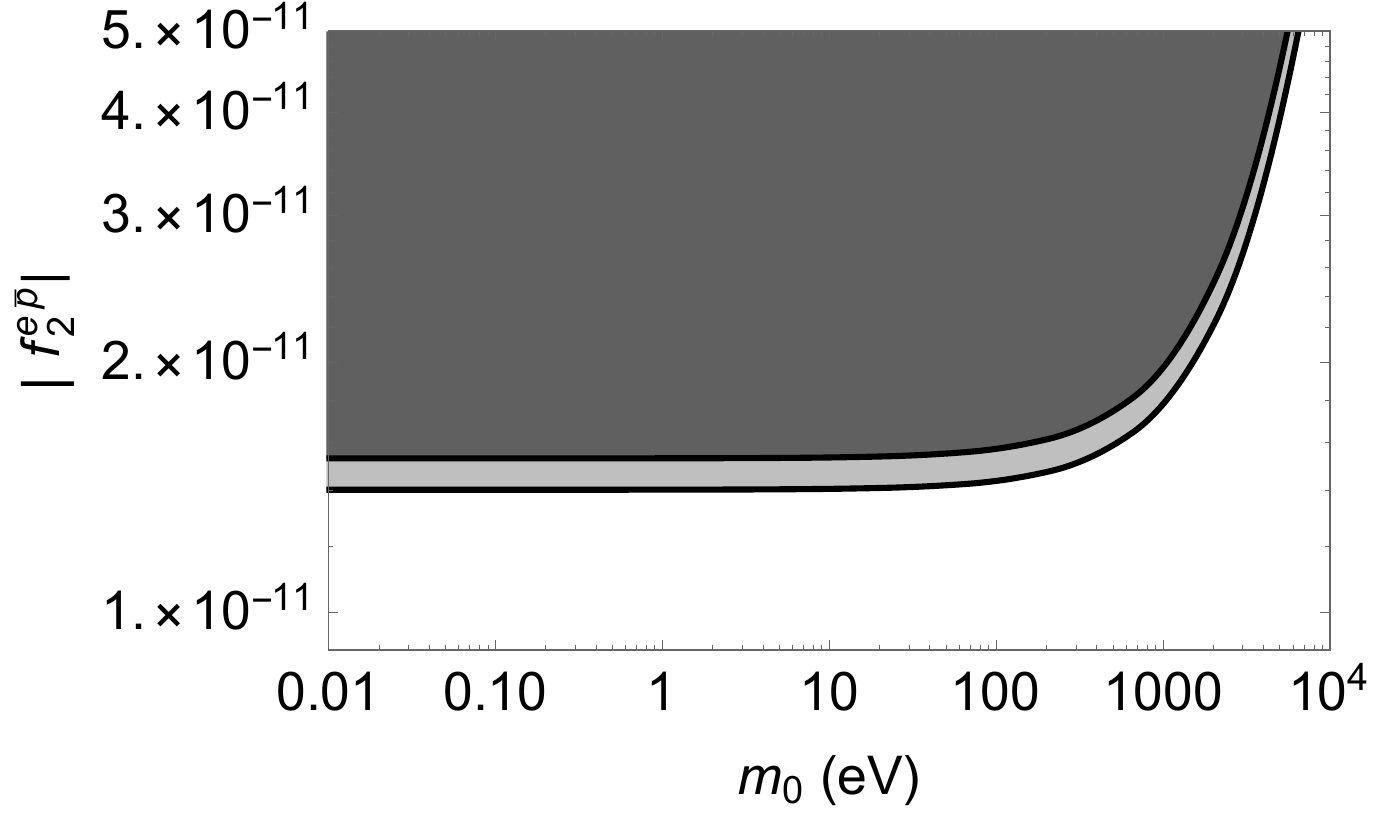}\\
\caption{\label{fig:check}Comparison of constraints (at the 90$\%$ confidence level) on the absolute value of the dimensionless coupling constant $f_2^{e\overline{p}}$ obtained with the wavefunction using effective charges coming from the variational method (lighter region) and the sensitivity check (darker region).}
\end{figure}

We are now ready to perform the last step, which is to use these effective charges and the wavefunction $\Psi$ to get constraints on the interaction constant $f^{e\overline{p}}_2$. We do this exactly as described in the main text and compare the results with the ones obtained using the variational method (see Fig. \ref{fig:v2}) in Fig. \ref{fig:check}. The resulting constraints agree to within several percent.

\section{Effective form of $V_{4+5}$ potential}\label{sec:v45}
In the limit of a massless boson ($\lambdabar\to\infty$), the potential $V_{4+5}$ takes the form
\begin{align} 
V_{4+5}=A_{4+5}^{e\overline{p}}  \textbf{s}_e \cdot \left[\left(\frac{m_e}{m_{\overline{p}}+m_e}\nabla_{\overline{p}}-\frac{m_{\overline{p}}}{m_{\overline{p}}+m_e}\nabla_e\right)\times\textbf{r},\frac{1}{r^3}\right]_{+},
\label{eqn:V4p5}
\end{align}
where $A_{4+5}^{e\overline{p}}$ includes a coupling parameter and other constants [cf. Eq.\ (\ref{eq:v5})]. We will show that for the considered antiprotonic helium states, the parts containing derivatives over the electron position may be neglected.

Let us focus on the expectation value $\langle \Psi | \textbf{V} |\Psi\rangle$, where $|\Psi\rangle$ here denotes the wavefunction of antiprotonic helium with the electron in the ground state and
\begin{align} 
\textbf{V}=\left[\nabla_e\times \textbf{r},\frac{1}{r^3}\right]_+.
\label{eqn:AV1}
\end{align}
Using $\textbf{r}=\textbf{r}_e-\textbf{r}_{\overline{p}}$, we may find that $\nabla_e\times \textbf{r}=-\textbf{r}\times\nabla_e$, so we expand
\begin{align} 
\left[\nabla_e\times \textbf{r},\frac{1}{r^3}\right]_+ \Psi=-\left[\textbf{r}\times\nabla_e,\frac{1}{r^3}\right]_+ \Psi\nonumber\\
=-\textbf{r}\times\nabla_e\left(\frac{1}{r^3}\Psi\right)-\frac{1}{r^3} \textbf{r}\times\nabla_e\Psi\nonumber\\
=-\Psi \textbf{r}\times\nabla_e\left(\frac{1}{r^3}\right)-\frac{2}{r^3} \textbf{r}\times\nabla_e\Psi.
\label{eqn:AV2}
\end{align}
The $i$-th component of the first term is
\begin{align} 
-\Psi\left(\textbf{r}\times\nabla_e \left(\frac{1}{r^3}\right)\right)_i=-\Psi\epsilon_{ijk} r^{j} \partial_e^k \left(\frac{1}{r^3}\right)\nonumber\\
=-\Psi\epsilon_{ijk} r^j \left(-\frac{3r^k_e}{r^5}\right)=\frac{3\Psi}{r^5}\epsilon_{ijk}(r^j_e-r^j_{\overline{p}}) r_e^k\nonumber\\
=-\left(\frac{3\Psi}{r^5}\textbf{r}_{\overline{p}}\times\textbf{r}_e \right)_i,
\label{eqn:AV3}
\end{align}
where $r^j$ is the $j$-th component of the vector $\textbf{r}$ and $\partial_e^k$ is the derivative with respect to the $k$-th direction in electron position space. To simplify the second term in Eq.\ (\ref{eqn:AV2}), let us recall that we assume that the electron is in the ground state, so the wavefunction $\Psi$ has the form
\begin{align} 
\Psi=\xi(\textbf{r}_{\overline{p}}) e^{-r_e/\lambda},
\label{eqn:AV4}
\end{align}
where $\xi$ is a function of the antiproton position only. We get
\begin{align} 
\textbf{r}\times\nabla_e\Psi=\xi(\textbf{r}_{\overline{p}})\textbf{r}\times\nabla_e \left(e^{-r_e/\lambda}\right),
\label{eqn:AV5}
\end{align}
and the $i$-th component is
\begin{align} 
\left(\textbf{r}\times\nabla_e\Psi\right)_i=\xi(\textbf{r}_{\overline{p}})\epsilon_{ijk} r^{j} \partial_e^k \left(e^{-r_e/\lambda}\right)\nonumber\\
=\xi(\textbf{r}_{\overline{p}})\epsilon_{ijk} r^{j} \left(-\frac{e^{-r_e/\lambda}}{\lambda r_e} r_e^k \right)=-\frac{\Psi}{\lambda r_e}\epsilon_{ijk}(r^j_e-r^j_{\overline{p}}) r_e^k\nonumber\\
=\left(\frac{\Psi}{\lambda r_e}\textbf{r}_{\overline{p}}\times\textbf{r}_e \right)_i.
\label{eqn:AV6}
\end{align}
Eventually, we obtain
\begin{align} 
\left[\nabla_e\times \textbf{r},\frac{1}{r^3}\right]_+ \Psi=&-\left(\frac{3}{r^5}+\frac{2}{\lambda r_e r^3}\right) (\textbf{r}_{\overline{p}}\times\textbf{r}_e)\Psi\nonumber\\
=&\left(\frac{3}{r^5}+\frac{2}{\lambda r_e r^3}\right) (\textbf{r}_e\times\textbf{r}_{\overline{p}})\Psi,
\label{eqn:AV7}
\end{align}
and the considered matrix element becomes
\begin{align} 
\langle \Psi | \textbf{V} |\Psi\rangle=\int_{\mathbb{R}^3} \int_{\mathbb{R}^3} \left(\frac{3}{r^5}+\frac{2}{\lambda r_e r^3}\right) (\textbf{r}_e\times\textbf{r}_{\overline{p}}) |\Psi|^2 ~ d^3\textbf{r}_{\overline{p}} ~ d^3\textbf{r}_e.
\label{eqn:AV8}
\end{align}

Let us now fix some value of $\textbf{r}_{\overline{p}}$ and perform the integration over the electron position. We define a vector function
\begin{align} 
\textbf{F}(\textbf{r}_{\overline{p}})=\int_{\mathbb{R}^3} \left(\frac{3}{r^5}+\frac{2}{\lambda r_e r^3}\right)  \textbf{r}_e e^{-2r_e/\lambda} ~ d^3\textbf{r}_e.
\label{eqn:AV9}
\end{align}
We choose a coordinate system $(r_e,\theta_e,\phi_e)$ with the zenithal direction parallel to $\textbf{r}_{\overline{p}}$. Then we may rewrite the function $\textbf{F}(\textbf{r}_{\overline{p}})$ as
\begin{align} 
\textbf{F}(\textbf{r}_{\overline{p}})=\int_0^\infty \int_0^\pi \int_0^{2\pi} \left(\frac{3}{r^5}+\frac{2}{\lambda r_e r^3}\right)\nonumber\\
\times r_e (\sin\theta_e \cos\phi_e,\sin\theta_e \sin\phi_e, \cos\theta_e) e^{-2r_e/\lambda} ~ dr_e ~ d\theta_e ~ d\phi_e,
\label{eqn:AV10}
\end{align}
where $r=(r_{\overline{p}}^2+r_e^2-2r_e r_{\overline{p}}\cos\theta_e)^{1/2}$. The integral over $\phi_e$ vanishes for the first and second components of this vector, so $\textbf{F}(\textbf{r}_{\overline{p}})$ is parallel to the zenithal direction, and hence parallel to $\textbf{r}_{\overline{p}}$.
This fact results in
\begin{align} 
\langle \Psi | \textbf{V} |\Psi\rangle=\int_{\mathbb{R}^3} (\textbf{F}(\textbf{r}_{\overline{p}})\times\textbf{r}_{\overline{p}}) |\xi(\textbf{r}_{\overline{p}})|^2 ~ d^3\textbf{r}_{\overline{p}}=\mathbf{0},
\label{eqn:AV11}
\end{align}
namely what we wanted to show.

\end{document}